\documentclass[journal,10pt,onecolumn]{IEEEtran}

\usepackage{cite}
\usepackage{amsmath,amssymb}
\usepackage{graphicx,bm,color,hyperref}

\usepackage{diagbox}
\usepackage{subcaption}
\usepackage{xcolor}

\usepackage{changes}

\newtheorem{theorem}{Theorem}

\newtheorem{lemma}[theorem]{Lemma}

\usepackage{tabularx}
\usepackage{booktabs}

\usepackage{mathrsfs}
\usepackage{dsfont}
\usepackage{mathtools}
\usepackage{bbm,bm}
\usepackage{mathdots}
 \usepackage{amssymb}
\usepackage{amsfonts}

\newcommand\ra{\rangle}
\newcommand\AME{\mathrm{AME}}
\newcommand\dH{d_H}\newcommand\dC{d}\newcommand\dloc{q}\newcommand{\trunc}[1]{_{\upharpoonright #1}} \DeclareMathOperator{\wt}{wt}\DeclareMathOperator{\lspan}{span}\DeclareMathOperator{\hiH}{\mathcal{H}}\DeclareMathOperator{\C}{\mathcal{C}}\DeclareMathOperator{\e}{\mathrm{e}}\DeclareMathOperator{\iu}{\mathrm{i}}\DeclareMathOperator{\Tr}{\mathrm{Tr}}\newcommand{\ket}[1]{\vert #1 \rangle}
\newcommand{\bra}[1]{\langle #1 \vert}
\newcommand{\ketbra}[2]{\vert #1 \rangle\langle #2\vert}
\newcommand{\braket}[2]{\langle #1 \vert #2 \rangle}
\newcommand{\ceil}[1]{\left\lceil #1 \right\rceil}
\newcommand{\floor}[1]{\left\lfloor #1 \right\rfloor}
\newcommand{\1}{\ensuremath{\mathbbm{1}}}
 
\newtheorem{conjecture}{Conjecture}
\newcommand{\Z}{\mathbb{Z}}
\renewcommand{\emph}{\textit}
\begin{document}

\title{Constructing optimal quantum error correcting codes from absolute maximally entangled states}

\author{\IEEEauthorblockN{Zahra Raissi\IEEEauthorrefmark{1}\IEEEauthorrefmark{2}, 
Christian Gogolin\IEEEauthorrefmark{2}, 
Arnau Riera\IEEEauthorrefmark{2},
and Antonio Ac\'{i}n\IEEEauthorrefmark{2}\IEEEauthorrefmark{3}}\\\medskip
  \IEEEauthorblockA{
    \IEEEauthorrefmark{1}Department of physics, Sharif University of Technology, Tehran, P.O. Box 111555-9161, Iran \\
    \IEEEauthorrefmark{2} ICFO-Institut de Ciencies Fotoniques, The Barcelona Institute of Science and Technology, Castelldefels (Barcelona), 08860, Spain\\
    \IEEEauthorrefmark{3}ICREA-Instituci\'o Catalana de Recerca i Estudis Avan\c cats, Lluis Companys 23, Barcelona, 08010, Spain}\vspace*{-4ex}
}

\maketitle

\begin{abstract}
 Absolutely maximally entangled (AME) states are pure multi-partite generalizations of the bipartite maximally entangled states with the property that all reduced states of at most half the system size are in the maximally mixed state.
AME states are of interest for multipartite teleportation and quantum secret sharing and have recently found new applications in the context of high-energy physics in toy models realizing the AdS/CFT-correspondence.
We work out in detail the connection between AME states of minimal support and classical maximum distance separable (MDS) error correcting codes and, in particular, provide explicit closed form expressions for AME states of $n$ parties with local dimension $\dloc$ a power of a prime for all $\dloc \geq n-1$.
Building on this, we construct a generalization of the Bell-basis consisting of AME states and develop a stabilizer formalism for AME states.
For every $\dloc \geq n-1$ prime we show how to construct QECCs that encode a logical qudit into a subspace spanned by AME states.
Under a conjecture for which we provide numerical evidence, this construction produces a family of quantum error correcting codes $[\![n,1,n/2]\!]_\dloc$ for $n$ even, saturating the quantum Singleton bound.
We show that our conjecture is equivalent to the existence of an operator whose support cannot be decreased by multiplying it with stabilizer products and explicitly construct the codes up to $n = 8$.
\end{abstract}

\vspace*{-3mm}
\section{Introduction}

A striking feature of quantum mechanics is \emph{entanglement} and the fact that having complete knowledge of the state of a system does not imply that one has complete knowledge of its subsystems. 
A paradigmatic example is an EPR state, in which a pure state of 2-qubits
has reduced density matrices on each half of the system that are completely mixed.
The family of states generalizing this property of EPR states to an arbitrary number of parties and local dimensions is the family of \emph{Absolutely maximally entangled} (AME) states.
AME states are pure quantum states of $n$-partite systems of local dimension $\dloc$ with the property that all reduced states (marginals) of at most half the system size are maximally mixed.

Just like EPR states, AME states are known to play an important role in quantum information processing when dealing with many parties.
They are useful for multipartite teleportation and in quantum secret sharing \cite{Helwig2012}.
AME states have also deep connections with apparently unrelated areas of mathematics such as combinatorial designs and structures \cite{Arnau},
classical error correcting codes \cite{Helwig2013}, 
and quantum error correcting codes (QECC) \cite{Scott}. 
Recently, they have gained new relevance as building blocks for holographic theories and in high energy physics.
There they allow for the construction of tensor network states that realize discrete instances of the AdS/CFT correspondence and holography \cite{Latorre,Pastawski,Harlow,Hayden}. 
AME states are special cases of the class of so-called $k$-uniform states for $k=\floor{n/2}$ \cite{Arnaud,Goyeneche2014}. 

At the same time it is still largely unknown for which $n$ and $\dloc$ AME states exist and how they can be constructed.
In the case of qubits for instance, it has been proven analytically that there are no AME states for $n=4$ and $n\ge 7$.
The non-existence in the cases $n=4$ and $n\geq 8$ was proven by finding a contradiction in a linear program \cite{Rains1999,Scott}.
Qubit AME states for $n=2,3$ were long known, a state for $n=5$ was found in \cite{Laflamme1996} and more recently such for $n=5,6$ were found numerically in \cite{Borras2007,Facchi2008,Facchi2010}.
Only very recently it was shown that there can not be a qubit AME state for the case $n=7$ \cite{Huber2016}.

A relevant class of AME states is formed by those states which, when written in the computational basis, have support on only $\dloc^{\floor{n/2}}$ basis states \cite{Bernal}. 
We call these states minimal support AME states. 
There is a direct correspondence between minimal support AME states and classical maximal distance separable (MDS) error correcting codes \cite{Arnau,Helwig2013}.
In this paper we work out the details of the minimal support AME-MDS correspondence and provide explicit constructions and closed form expressions for AME states for arbitrary $n$ and for all $\dloc \geq n-1$.
Further, from a single AME state, we show how to produce an orthonormal basis of AME states.
Based on our construction of minimal support AME states, we introduce stabilizer operators and conjecture the existence of a family of QECC whose code spaces are spanned by AME states.
We show that our conjecture is equivalent to the existence of a Pauli string that is incompressible in the sense that its weight cannot be decreased by multiplying it with stabilizer products and connect this with a feature of the joint weight enumerators of certain MDS codes.
Further we construct such codes for all $n$ up to $n=8$ by finding several suitable incompressible operators.
In these QECCs, a logical qudit is encoded in a $\dloc$-dimensional subspace spanned by AME states of $n$ parties.
Our proposal has a very clear physical motivation and nicely complements other constructions of non-binary QECC (see the end of Section~\ref{sec:qerrfromamestates} for a comparison).
In particular our construction is very explicit and works with a smaller local dimension $\dloc$ given $n$ than previous codes with similar code parameters.

The paper is organized as follows: 
In Section~\ref{sec:amestates} we review basic properties of AME states.
In Section~\ref{sec:ame-states-from-mds-codes} we focus on the correspondence between the AME states of minimal support and MDS codes.
In Section~\ref{sec:constructionofgeneratormatrices} constructions of linear MDS codes are introduced and a systematic construction of AME states of minimal support is presented.
Then, in Section~\ref{sec:amebasis}, we show how to construct an orthonormal basis of AME states from any given AME state.
Finally, in Sections~\ref{sec:stabilizers} and \ref{sec:qerrfromamestates}, we develop a stabilizer formalism for AME states of minimal support constructed from an MDS code and discuss the construction of optimal AME based QECC codes that saturate the quantum Singleton bound.

\section{AME states}
\label{sec:amestates}
We begin with introducing some general notation:
Given a real number $r \in \mathbbm{R}$ we denote by floor $r$, $\floor{r}$, the largest integer not larger than $r$ and by ceiling $r$, $\ceil{r}$, the smallest integer not smaller than $r$.
For any $n \in \Z^+$ we denote by $[n] \coloneqq (0,\dots,n-1)$ the range from $0$ to $n-1$.
Let $\hiH(n,\dloc) \coloneqq \mathbb{C}_\dloc^{\otimes n}$ be the Hilbert space of $n$ distinguishable $\dloc$ level quantum systems (also called qudits).
For any sequence $j_1,\dots,j_n \in [\dloc]^{n}$ we denote the corresponding vector by $\vec j \coloneqq (j_1,\dots,j_n)$, its length by $|\vec j| = n$, and write $\ket{\vec j} \coloneqq \ket{j_1,\dots,j_n} \coloneqq \ket{j_1} \otimes \dots \otimes \ket{j_n}$ for the associated computational basis state in $\hiH(n,\dloc)$.
For any sequence $j_1,\dots,j_n$ and subset $S \subset [n]$ of indices, we denote the truncation of $\vec j$ to the index set $S$ by $\vec{j}\trunc{S} \coloneqq (j_l)_{l\in S}$.
For instance, given the vector $\vec j=(6,4,3,4,5)$ and the subset $S = \{1,2,5\}$ we have $\vec{j}\trunc{S}=(6,4,5)$.
Further, we denote the set of AME states in $\hiH(n,\dloc)$ by $\AME(n,\dloc)$.

Absolutely maximally entangled states are those pure states whose reduced states on at most half of the systems are all maximally mixed, concretely
\begin{equation}
  \begin{split}
    \AME(n,\dloc) \coloneqq \qquad& \\
    \{\ket\psi\in \hiH(n,\dloc) \colon &\forall S \subset \{1,\ldots,n\}\ |S|\geq \ceil{n/2}\\
    &\implies \Tr_S \ketbra\psi\psi \propto \1 \} \ .
  \end{split}
\end{equation}
For a suitable choice of the computational basis, any AME state can be written as 
\begin{equation} \label{eq:ame-in-computational-basis}
  \AME(n,\dloc) \ni \ket\Psi = \sum_{j_1,\dots,j_n=0}^{\dloc-1} c_{j_1,\dots,j_n} \ket{j_1\dots,j_n}\, ,
\end{equation}
where the coefficients $c_{j_1\dots,j_n}$ can be regarded as a tensor of $n$ indices with  the property of being \emph{multi-unitary} \cite{Arnau} or \emph{perfect} \cite{Pastawski}.
A tensor $c$ is called perfect if for any bipartition of its indices into a set $S$ and complementary set $S^c$ with $|S| \le |S^c|$, the resulting matrix $C \coloneqq c_{\vec{j}\trunc{S},\vec{j}\trunc{S^c}}$ is an isometry, i.e., $C\,C^\dagger=\1$.
Using this matrix $C$, Eq.~(\ref{eq:ame-in-computational-basis}) can be rewritten as
\begin{equation} \label{eq:ame-in-terms-of-a-perfect-tensor}
\ket\Psi =\sum_{\vec{l} \in [\dloc]^{|S|}} \ket{\vec{l}} \otimes C\,\ket{\vec{l}} \, .
\end{equation}
Note that the states $C \ket{\vec{l}} \in \hiH(|S^c|,\dloc)$ are in general not product states.

AME states can be classified according to the number of terms in their expansion in the computational basis \cite{Arnau}.
The maximum number of terms of a state in $\AME(n,\dloc)$ is obviously $\dloc^{n}$ and such AME states are called \emph{maximal support} AME states.
On the contrary, the minimal number of terms for which the condition of maximally mixed marginals can still be fulfilled is the dimension of the largest sub-system on which the state is required to still be maximally mixed, namely $\dloc^{\floor{n/2}}$.
AME states with this many terms are called \emph{minimal support} AME states and will be at the center of attention in this work.

\section{Correspondence between minimal support AME states and MDS codes}
\label{sec:ame-states-from-mds-codes}
There is a direct correspondence between minimal support AME states and classical MDS codes \cite{Helwig2013}.
We first describe how an MDS code can be obtained from any minimal support AME state, then define more generally what an MDS code is, and finally show how any MDS code that encodes $\floor{n/2}$ dits ($\dloc$ level classical systems) into $n$ dits allows for the construction of minimal support AME states in $\AME(n,\dloc)$.

Being an AME state of minimal support puts strong constraints on the coefficients $c_{j_1,\dots,j_n}$ in the expansion in Eq.~\eqref{eq:ame-in-computational-basis}.
Let the Hamming distance between two sequences $j_1,\dots,j_n$ and $k_1,\dots,k_n$ be the number of sub-indices $l$ for which $j_l \neq k_l$.
For example, the Hamming distance between the sequences $0030$ and $0130$ is $1$, since they differ in exactly one position.
Then, $\ket\Psi$ can only be a minimal support AME state if $|c_{j_1\dots,j_n}| \in \{0,1/\sqrt{\dloc^{\floor{n/2}}}\}$ and if all sequences $j_1\dots,j_n$ for which $|c_{j_1\dots,j_n}| \neq 0$ have pairwise Hamming distance at least $\ceil{n/2}+1$ \cite{Arnau}.
To see this, let us consider a bipartiton $S \cup S^c = \{1,\dots,n\}$ and look at $C = c_{\vec{j}\trunc{S},\vec{j}\trunc{S^c}}$ as a linear map from the space $\hiH(|S|,\dloc)$ to $\hiH(|S^c|,\dloc)$.
As $\ket\psi$ is of minimal support, the states $C\ket{\vec{j}\trunc{S}}$ are computational basis states and hence every column of the matrix $C$ contains only a single non-zero element.
Now consider the case $|S| = \floor{n/2}$.
Then, $C$ associates to any sequence $\vec l$ of length $\floor{n/2}$ a sequence $\vec m$ of length $\ceil{n/2}$, namely the one for which $C_{\vec{l},\vec{m}} \neq 0$.
As the AME state is minimal support, there are precisely $\dloc^{\floor{n/2}}$ such sequences of length $\ceil{n/2}$.
Consider now the set of sequences that is obtained by concatenating any sequence of length $\floor{n/2}$ with the associated sequence of length $\ceil{n/2}$, i.e., the set $\{\vec{l} \circ \vec {m} : \vec{l} \in [\dloc]^{\floor{n/2}} \wedge C_{\vec{l},\vec{m}} \neq 0\}$.
This set of sequences, with large mutual Hamming distance, yields a so-called maximal distance separable (MDS) classical error correcting code \cite[Chapter~11]{MacWilliams},\cite{Singleton,Roth,Maruta90}.

Given integers $n,k,\dloc$ with $n \geq k$, an error correcting code is an injective mapping from the space $[\dloc]^{k}$ of $k$ dit messages to a subset of the set of $n$ dit code words.
If $n > k$, protection against errors on some of the dits can be achieved.
The protection depends on the Hamming distance between the code words.
In the language of coding theory a $[n,k,\dH]_\dloc$-code is an error correcting code that works with $q$-level dits and encodes messages of length $k$ into code words of length $n$, all having pairwise Hamming distance at least $\dH$.
Such a code protects against errors on any subset of a most $t = \floor{(\dH-1)/2}$ many dits \cite[chapter~1]{MacWilliams},\cite{Reed}.
The code obtainable from a minimal support AME state in $\AME(n,\dloc)$ by the construction described above is thus a $[n,\floor{n/2},\ceil{n/2}+1]_\dloc$-code.

A code with a given $n$ and $k$ is called an MDS code if the minimal Hamming distance between any two code words satisfies the Singleton bound.
The Singleton bound is a fundamental result from coding theory that bounds the maximally achievable minimal Hamming distance between any two code words in a code.
By comparing the combined volume of Hamming distance balls of radius $\dH$ around the $\dloc^k$ code words with the total available volume in the set of sequences of length $n$ one can show that any code has to satisfy \cite{Singleton}
\begin{equation} \label{eq:singletonbound}
  \dH \leq n - k + 1 .
\end{equation}
One directly verifies that the $[n,\floor{n/2},\ceil{n/2}+1]_\dloc$-code constructed above verifies this bound with equality for all $n$.
Thus from any minimal support AME state a classical MDS code can be constructed.

Conversely, we now describe how from any suitable linear MDS code a corresponding AME state can be constructed.
In order to do this we first need to introduce a notion of linear independence that is suitable for sequences over a finite set of elements.
This brings us to the theory of finite fields \cite[chapter~3,4]{MacWilliams}.
A finite field (or Galois field) is a finite set of elements that is closed under addition, subtraction, (commutative) multiplication and division (excluding division by zero).
For every prime number $p$ and every natural number $m$ there exists exactly one finite field $GF(p^m)$ of cardinality (also called order) $p^m$.
For every prime $p$ the finite field $GF(p)$ is equal to the integers modulo $p$.
The non-prime finite fields $GF(p^m)$ can be explicitly constructed as follows:
Let $GF(p)[x]$ be the set of polynomials in $x$ over $GF(p)$, that is, the polynomials whose coefficients, variable $x$, addition and multiplication are those from $GF(p)$.
Choose a polynomial $P$ over $GF(p)$ of order $m$ that is irreducible with respect to that field.
Irreducible here means that $P$ can not be written as the product of two non-constant polynomials in $GF(p)[x]$.
The existence of such a polynomial $P$ is always guaranteed \cite[chapter~3]{MacWilliams}.
$GF(p^m)$ is then the quotient ring $GF(p^m) = GF(p)[x]/P$, that is, $GF(p^m)$ is the set of polynomials of order less than $m$ with the standard addition and subtraction of polynomials over $GF(p)$ and the result of multiplication is the remainder after Euclidean division by $P$.

The encoding map of a linear code is a linear map from the space of messages to the space of codewords, called a generator matrix $G_{k \times n}$.
The encoded version of an arbitrary message can be obtained by splitting the message up into blocks of length $k$ and multiplying the corresponding row vectors from the right with the $k \times n$ generator matrix, thereby yielding the corresponding code word.
Multiplication and addition are thereby to be performed in a finite field whose cardinality is at least as large as that of the message alphabet.

It turns out that the generator matrix $G_{k \times n}$ of any $[n,k,\dH]$-code over a finite field $GF(p^m)$ can always be written in the standard form \cite[chapter~1]{MacWilliams}
\begin{equation}
  G_{k\times n}=[\1_k|A] \label{generator}
\end{equation}
where $\1_k$ is a $k \times k$ identity matrix and $A \in GF(p^m)^{k \times (n-k)}$.
This standard form will be useful several times in the following.

Every linear code $\mathscr{C}$ has a dual code $\mathscr{C}^{\perp}$, that is the code whose code words are orthogonal to all the codewords of the original code with respect to the standard Euclidean inner product of the finite field.
The generator matrix $H_{n-k \times n}$ of the dual code is the so-called parity check matrix of the original code.
It satisfies $G_{k \times n}\,(H_{n-k \times n})^T = 0$ and if $G_{k \times n}$ is given in standard form, then $H_{n-k \times n} = [-A^T|\1_{n-k}]$.
The matrix $H_{n-k \times n}$ is called parity check matrix, because for any code word $\vec{c} \in \mathscr{C}$ of the original code $H_{n-k \times n}\,\vec{c}^{\ T} = 0$, so it can be used to check whether a string is a code word or not.

Given the generator matrix $G_{\floor{n/2} \times n}$, or alternatively the $A$ matrix, of a suitable MDS code with $k = \floor{n/2}$ over a finite field of cardinality $\dloc$ (equal to a power of a prime), it is straightforward to construct an AME state in $\AME(n,\dloc)$.
As the Hamming distance of the MDS code is $\dH = \ceil{n/2}+1$, for any two different $\vec{v},\vec{w} \in [q]^{\floor{n/2}}$ the states $\ket{\vec{v}\,G_{\floor{n/2} \times n}}$ and $\ket{\vec{w}\,G_{\floor{n/2} \times n}}$ are orthogonal on all subsystems of size at least $\ceil{n/2}$.
The state 
\begin{equation}
  \ket\Psi = \sum_{\vec{v} \in [\dloc]^{\floor{n/2}}} \ket{\vec{v}\,G_{k \times n}} = \sum_{\vec{v} \in [\dloc]^{\floor{n/2}}} \ket{\vec{v} ,  \vec{v}\,A}. \label{amemingenerator}
\end{equation} 
is hence a minimal support AME state in $\AME(n,\dloc)$.

AME states can hence be constructed whenever a suitable $G_{k \times n}$ or $A$ matrix is known. 
The first examples of such $A$ matrices were presented by Singleton \cite{Singleton} for the cases $\dloc=p^m=5$ and $\dloc=p^m=7$ and later a general construction was found in \cite{Roth, Seroussi}.
We will come back to explaining how suitable matrices $A$ and $G_{k \times n}$ can be constructed in the next section.
First, as an example, we go through the construction of a minimal support state in $\AME(6,5)$. 
In this example the local dimension $\dloc=5$ is prime so that the finite field $GF(5)$ is simply the set $\{0,1,2,3,4\}$ with the standard arithmetic modulo $5$.
The number of free indices in the closed form expression of the AME state with minimal support is $k=\floor{n/2}=3$, so we can write $\vec{v}=(i,j,l)$.
A suitable generator matrix of a $[6,3,4]_5$ MDS code in standard form is
\begin{equation}
G_{3 \times 6} =\left[  \begin{array}{ccc|ccc}
1 & 0 & 0 & 1 & 1 & 1 \\
0 & 1 & 0 & 1 & 2 & 3 \\
0 & 0 & 1 & 1 & 3 & 4  \\
 \end{array} \right] .
\end{equation}
It yields the following closed form expression for a minimal support state in $\AME (6,5)$:
\begin{equation}
  \begin{split}
    &\AME(6,5) \ni \ket\Psi = \sum_{\vec v \in GF(5)^3} \ket{\vec v \,G} \\
    &= \sum_{i,j,l=0}^4 |i,j,l,i+j+l,i+2j+3l,i+3j+4l\ra \label{AME6,5}    
  \end{split}
\end{equation}
For an example with a non-prime finite field see Appendix~\ref{app:construction-Singleton-AMEstate}.

\section{Explicit construction of generator matrices for MDS codes and AME states}
\label{sec:constructionofgeneratormatrices}
We now show explicitly how generator matrices of MDS codes and hence minimal support AME states can be constructed and how closed formulas, reminiscent of the example in the end of the last Section, can be obtained for all $n$.
To do this, we first discuss the properties of the generator matrices of MDS codes in more detail.

Coming back to the standard form of the generator matrices of linear codes
$G_{k\times n}=[\1_k|A]$, we can readily see that a code can only be an MDS code if all entries of the $A$ matrix are non-zero and that $\dH=n-k+1$ is the optimal achievable Hamming distance between all code words.
If $A$ had a zero somewhere, there would be a code word with less than $n-k+1$ non-zero symbols and which hence would have Hamming distance less than $n-k+1$ to the all zero code word (which is a valid code word in any linear code).

In fact, a linear code with a given $A$ matrix is an MDS code if and only if every square submatrix of $A$ is nonsingular \cite[Chapter~11]{MacWilliams}, \cite{Singleton}.
To show this we first need to prove the following:
Every square submatrix of $A$ is nonsingular if and only if any subset of at most $k$ of the column vectors of $G_{k\times n}=[\1_k|A]$ is linearly independent. 
First note that it is enough to show this for subsets of size exactly $k$.
Let now $K$ be the square matrix of any given set of $k$ column vectors of $G_{k\times n}$.
These vectors are linearly independent if and only if the determinant $\det(K)$ is non-zero.
By shuffling all the columns that came from the $\1_k$ part of $G_{n\times k}$ to the left and then using Laplace's expansion of $\det(K)$ in terms of the determinants of minors, one realizes that $\det(K)$ is (up to possibly a sign) equal to the determinant of a square sub-matrix of $A$, and hence non-zero.
This is true for all sub-sets of at most $k$ columns only if all sub-matrices of $A$ are non-singular.

As we are looking at linear codes, and hence any linear combination of code words is again a valid code word, to find the minimal distance between any two code words it is sufficient to find the code word with the minimal Hamming distance from the all zero code word.
This however is exactly $n$ minus the number of zeros that we can generate by taking linear combinations of the rows of $G_{k\times n}$.
This, due to the linear independence of the subsets of columns, being at most $k-1$, implies that the achieved Hamming distance is exactly $\dH = n -k + 1$, saturating the Singleton bound.

To construct suitable $A$ matrices, we now introduce the concept of so-called \emph{Singleton arrays}.
Any finite field $GF(\dloc)$, with $\dloc$ a power of a prime, contains at least one primitive element \cite[chapter~4]{MacWilliams}.
An element $\gamma \in GF(\dloc)$ is called primitive if all the non-zero elements of $GF(\dloc)$ can be written as some integer power of $\gamma$.
Given any such primitive element $\gamma$, the Singleton array of size $q$ is defined to be
\begin{equation}
S_\dloc \coloneqq \begin{array}{ccccccc}
1 & 1 & 1 & \ldots & 1 & 1 & 1\\
1 & a_1 & a_2 & \ldots & a_{\dloc-3} & a_{\dloc-2} &  \\
1 & a_2 & a_3  & \ldots & a_{\dloc-2} &   &  \\
\vdots & \vdots & \vdots & \iddots &  &   &  \\
1 & a_{\dloc-3} & a_{\dloc-2} &   &   &   &    \\
1 & a_{\dloc-2} &  &   &   &   &   \\
1 &   &  &   &   &   &    \\ 
 \end{array}, \label{sq}
\end{equation}
with
\begin{equation}
  a_i \coloneqq \frac{1}{1-\gamma^i} .
\end{equation}
The Singleton array is a special case of a more general construction known as a Cauchy matrix \cite[chapter~11]{MacWilliams}.
Every submatrix of a Cauchy matrix is again a Cauchy matrix and an explicit formula for the determinant of any Cauchy matrix is known, which in particular shows that it is non-zero.
The Singleton array $S_\dloc$ thus has the sought after property that all its square sub matrices are non-singular \cite{Roth,Maruta90}.
By taking rectangular sub-matrices of $S_q$, it is hence possible to construct generator matrices $G_{k\times n}=[\1_k | A_{k,n-k}]$ of MDS codes and thereby minimal support AME states.
All one has to do is to take a power of a prime $\dloc$ sufficiently large such that $S_\dloc$ contains a sub-matrix of size at least $\floor{(\dloc+1)/2} \times \ceil{(\dloc+1)/2}$, and then take this as the matrix $A$ in Eq.~\eqref{amemingenerator}.
We provide a Mathematica notebook for the explicit construction of Singleton arrays under \cite{amegithub}. 

One straightforwardly verifies that $S_\dloc$ contains such a sufficiently large sub-matrix whenever $\dloc\geq n -1$.
Further, if $\dloc$ is even, the element $a_1$ can be appended to the third and the $(\dloc-1)$-st rows of $S_q$, without creating singular submatrices \cite[chapter~11]{MacWilliams},\cite{Roth}.
For $\dloc=4$ this increases the size of the largest square sub-matrix, as the extended Singleton array $S_4'$ has the form
\begin{equation} \label{eq:extendedlingletionarray}
S_4' = \begin{array}{cccc}
1 & 1 & 1 & 1\\
1 & a_1 & a_2 &   \\
1 & a_2 & a_1  &   \\
1 &   &  &  \\ 
 \end{array}.
\end{equation}
This yields an $A$ matrix of size $3\times 3$ for $\dloc=4$, giving a closed form formula for a minimal support AME state in $AME(6,4)$. 

As an example, let us consider the case $\dloc=5$.
Taking $\gamma=3$, which is a primitive element in $GF(5)=\{0,1,2,3,4\} \mod(5)$, we find
\begin{align}
a_1 &= \frac{1}{1-3}= \frac{1}{3} = 2 \\
a_2 &= \frac{1}{1-9}=\frac{1}{2}=3 \\
a_3 &= \frac{1}{1-27}=\frac{1}{4}=4
\end{align}
and obtain
\begin{equation}
S_5 = \begin{array}{ccccc}1 & 1 & 1 & 1 &1 \\ 1 & 2 & 3 & 4 & \\ 1 & 3 & 4 &   & \\ 1 & 4 &  &  & \\ 1 &   &   &   &  \end{array}.
\end{equation}
The biggest submatrix has size $3\times 3$.
Hence, taking
\begin{equation}
A_{3\times 3}=\left[\begin{array}{ccc}
1 & 1 & 1 \\
1 & 2 & 3 \\
1 & 3 & 4  \\
 \end{array} \right]
\end{equation}
we can construct a $[6,3,4]_5$-code, which is an MDS code, and the resulting AME state is precisely the one given in Eq.~\eqref{AME6,5}.
We list Singleton arrays for various finite fields in Appendix~\ref{app:construction-Singleton-AMEstate} and provide a Mathematica notebook for their construction \cite{amegithub}. 

\section{Basis of AME states}
\label{sec:amebasis}
The Bell basis of the Hilbert space of 2 qubits is an orthonormal basis of maximally entangled states.
In what follows, we show how, starting from a single AME state in $\hiH(n,\dloc)$, a complete orthonormal basis of AME states for $\hiH(n,\dloc)$, a \emph{AME basis}, can be constructed.

First we introduce operators $X$ and $Z$ that generalize the Pauli $\sigma_X$ and $\sigma_Z$ operators in Hilbert spaces of dimension $\dloc \geq 2$, defined via their action on the computational basis states $\ket{j}$
\begin{align} 
X\ket{j}&=\ket{j+1 \mod \dloc} \label{eq:defX}\\
Z\ket{j}&=\omega^j\,\ket{j}\, , \label{eq:defZ}
\end{align}
with $\omega \coloneqq \e^{\iu 2 \pi/\dloc}$ the $\dloc$-th root of unity.
Note that $X$ and $Z$ are unitary, traceless, and that $X^\dloc = Z^\dloc = \1$ and that for $a,b \in [\dloc]$ it holds that $\Tr(Z^a\,X^b) = \delta_{a,0}\,\delta_{b,0}$ and $Z\,X = \omega\,X\,Z$.
We call operators that are tensor products of the Pauli operators \emph{Pauli strings}.
As a side remark, note that only for $\dloc$ prime are the integers modulo $\dloc$ equal to the finite field $GF(q)$.
In all other cases, the algebraic structure of $X$ and $Z$ as defined above does not correspond to that of the respective finite field (if it even exists).
This however is irrelevant for this section.
The following works for arbitrary $\dloc$ not necessarily prime or a power of a prime.
Only the properties of $X$ and $Z$ discussed above are used.

For every $\vec a$ we define the operator
\begin{align}
\label{eq:Moperatordef}
& M(\vec a) \coloneqq \\
& (\underbrace{\1  \dots  \1}_{\ceil{n/2}} \otimes Z^{a_1}  \dots  Z^{a_{\floor{n/2}}})\, (\underbrace{\1  \dots  \1}_{\floor{n/2}} \otimes X^{a_{\floor{n/2}+1}}  \ldots  X^{a_n}) \, . \nonumber
\end{align}
Note that, for $n$ even, the potential number of $X$'s and that of $Z$'s are equal, namely $n/2$.
In contrast, if $n$ is odd, the maximal number of $X$'s is one larger than the maximal number of $Z$'s.

We now use this family of operators to construct, given any AME state, a complete orthonormal basis of AME states:
\begin{lemma}
  Given a Hilbert space $\hiH(n,\dloc)$ of $n$ parties with local dimension $\dloc$ with at least one AME state $\ket{\Psi} \in \hiH(n,\dloc)$, then the $\dloc^n$ states 
  \begin{equation}
    \ket{\Psi_{\vec{a}}} \coloneqq M(\vec{a})\,\ket{\Psi}
  \end{equation}
  with $\vec a \in [q]^n$ form a complete orthonormal basis of AME states of $\hiH(n,\dloc)$.
\end{lemma}
\begin{IEEEproof}
  First, all the $\ket{\Psi_{\vec{a}}}$ are AME states, as acting with local unitaries on $\ket\Psi$ does not change the entanglement properties.
  It remains to show orthonormality, i.e., that
  \begin{equation}
    \bra{\Psi} M(\vec a)^\dagger\, M(\vec b) \ket{\Psi} = \prod_i \delta_{a_i,b_i} \, .
  \end{equation}
  To show this we use that according to Eq.~\eqref{eq:ame-in-terms-of-a-perfect-tensor} any AME state $\ket{\Psi}$ can be written as 
  \begin{equation}
    \ket\Psi =\sum_{\vec{l} \in [\dloc]^{|S|}} \ket{\vec{l}} \otimes C\,\ket{\vec{l}} \, ,
  \end{equation}
  with $|S| = \floor{n/2}$ and $C$ an isometry, i.e., $C\,C^\dagger = \1$.
  Thus
  \begin{align}
 \bra{\Psi} M(\vec a)^\dagger & M(\vec b) \ket{\Psi} \cr
& =\sum_{\vec{l},\vec{m} \in [\dloc]^{|S|}}  \braket{\vec{l}}{\vec{m}}\, \bra{\vec{l}}\,\C^\dagger\,M(\vec a)^\dagger\,M(\vec b)\, C\,\ket{\vec{m}}\\
    &= \sum_{\vec{l} \in [\dloc]^{|S|}}  \bra{\vec{l}}\,\C^\dagger\,M(\vec a)^\dagger\,M(\vec b)\, C\,\ket{\vec{l}}\\
    &= \Tr(M(\vec a)^\dagger\,M(\vec b)\, C\,C^\dagger) \\
    &= \prod_i \delta_{a_i,b_i} \, ,
  \end{align}  
  where we have used the orthonormality of the computational basis states, the cyclicity of the trace and that $\Tr(Z^a\,X^b) = \delta_{a,0}\,\delta_{b,0}$.
\end{IEEEproof}

\section{Stabalizer operators for AME states}
\label{sec:stabilizers}
In the theory of quantum error correcting codes \cite{Gottesman,Gottesman-thesis} stabilisers are a useful tool to construct and analyse codes.
A quantum error correcting code distinguishes a subspace (code space) of the Hilbert space of a physical system as the space of admissible code states, that is, quantum states of the system that are in a one to one correspondence (via the encoding and decoding maps) with encoded messages.
For the code to be useful, the code space must be chosen such that the expected errors never map state from the code space to a state that could also have been produced by a different error from a different code state (this would introduce an unrecoverable error) but always take the state out of the code space in a away such that a subsequent correction can bring the system back into its original state.
It is natural to consider code spaces that are spanned by computational basis states.
The stabiliser (group) of such a code space is the abelian sub-group of the (generalized) Pauli group that leaves every element from the code space invariant.
Conversely every abelian sub-group of the (generalized) Pauli group that does not contain $-\1$ has a non-trivial subspace spanned by computational basis stats that is left invariant \cite{Gottesman,Gottesman-thesis}.

In an analogous fashion, we can construct a set of Pauli strings that generate a stabilizer group that stabilizes a given individual AME state.
In the next section we will use this generating set to construct a stabilizer group for a subspace spanned by $\dloc$ many orthonormal AME states.
The construction we present only works for AME states constructed from an MDS code as described in Section~\ref{sec:ame-states-from-mds-codes} and can hence only work for $\dloc$ a power of a prime.
For the sake of simplicity we further restrict from now on to the case $\dloc$ prime, for which the algebraic structure of the $X$ and $Z$ operators defined in \eqref{eq:defX} and \eqref{eq:defZ} coincides with that of the finite field $GF(\dloc)$.
For $\dloc$ a power of a prime, a much more elaborate construction based on the (discrete) Heisenberg-Weyl group \cite{Weyl,Wootters,Bandyopadhyay,Grassl2004,Durt2004} would have to be employed.

Remember that, given a generator matrix $G_{k\times n}$, the corresponding AME state takes the form (recall Eq.~\eqref{amemingenerator})
\begin{equation} 
  \ket\Psi = \sum_{\vec{v} \in [\dloc]^{\floor{n/2}}} \ket{\vec{v}\,G_{\floor{n/2} \times n}} .
\end{equation}
Denote the matrix elements of $G_{\floor{n/2}\times n}$ by $g_{l,m}$ and that of the code's parity check matrix $H_{\ceil{n/2} \times n}$ by $h_{l,m}$.
For $\dloc$ prime, the state $\ket\Psi$ is then the plus one eigenstate of the following $n$ stabilizer operators:
\begin{equation}\label{eq:psi-stabilizers}
  s^\Psi_l \coloneqq 
  \begin{cases}
    \bigotimes_{m=1}^{n} X^{g_{l,m}} & 1 \leq l \leq \floor{n/2} \\
    \bigotimes_{m=1}^{n} Z^{h_{l,m}} & \floor{n/2} < l \leq n
  \end{cases} .
\end{equation}
The first $\floor{n/2}$ stabilizers, involving the $X$ operators, permute the computational basis states in the decomposition of $\ket\Psi$ and hence leave it invariant.
The second set of $\ceil{n/2}$ stabilizers, that involve the $Z$ operators, also leave $\ket\Psi$ invariant as
\begin{align} 
  s^\Psi_l\,\ket\Psi = \sum_{\vec{v} \in [\dloc]^{\floor{n/2}}} \omega^{H_{\ceil{n/2} \times n}\,(G_{\floor{n/2} \times n})^T\,\vec{v}}\,\ket{\vec{v}\,G_{\floor{n/2} \times n}} = \ket\Psi ,
\end{align}
because $H_{\ceil{n/2} \times n}\,(G_{\floor{n/2} \times n})^T = 0$.

\section{Quantum error correcting codes from AME states}
\label{sec:qerrfromamestates}
In this section we show that the AME states of minimal support constructed from MDS codes allow to construct quantum error correcting codes (QECC).
Our construction is comparably simply, very explicit, physically motivated, and works with a smaller local dimension $\dloc$ given $n$ than previous codes with similar code parameters.

A subspace $\mathcal{C}$ spanned by a set $\{\ket{\psi_m}\}_{m \in [\dloc^k]}$ of orthonormal states is a $[\![n,k,\dC]\!]_\dloc$ a QECC, i.e., a code that encodes $k$ logical qudits (quantum systems of dimension $\dloc$) into $n$ physical qudits, if it obeys the Knill-Laflamme conditions \cite{Knill,Gottesman-thesis}
\begin{equation}
  \forall m,m' \in [\dloc^k] \colon\ \bra{\psi_m}E^\dagger F\ket{\psi_{m'}}=f(E^{\dagger}F)\,\delta_{m,m'}
\end{equation}
for all $E,F$ with $\wt(E^{\dagger}F) \leq \dC$.
Thereby $\wt$ is the \emph{weight} of an operator, defined to be the number of sites on which it acts non-trivially.
The parameter $\dC$ is the distance of the code, which is the minimal number of single-qudit operations that are needed to create a non-zero overlap between any two orthogonal states from the code state space $\mathcal{C}$, i.e., 
\begin{equation}
  \dC \coloneqq \min_{\ket{\phi},\ket{\phi'}\in \mathcal{C}, W} \{\wt(W) \colon 
  \bra{\phi}W\ket{\phi'}\ne 0 \land  \braket{\phi}{\phi'}= 0\}\, .
\end{equation}
Such a code can correct errors that act non-trivially on up to $t\coloneqq\floor{(\dC-1)/2}$ physical qudits.

The quantum Singleton bound \cite{Gottesman-thesis,Knill, Cerf} states that for any QECC
\begin{equation} \label{eq:quantumsingletonbound}
  \dC \leq \frac{n-k}{2}+1 .
\end{equation}
Comparing with the classical Singleton bound given in Eq.~\eqref{eq:singletonbound}, we see that, to reach a given code distance in the quantum case, $n-k$ must be twice as large as the necessary value to reach the same Hamming distance in the classical case.
The proof of the quantum Singleton bound is based on the no-cloning theorem and it is known that binary codes, that is codes for qubits $\dloc = 2$, can not achieve it for large $n$.

The code space of the QECC that we are going to construct will be spanned by AME states generated by acting with a Pauli string $M$ onto a given minimal support AME state $\ket{\Psi}$ constructed from an MDS code.
Let us first introduce the notion of different \emph{realizations} of such a Pauli string $M$.
Recall first that AME states generated by Eq.~\eqref{amemingenerator} are stabilized by a set of $\dloc^n$ Pauli strings, the elements of the stabilizer group, denoted by
\begin{equation}
  S(\alpha_1,\ldots,\alpha_n) \coloneqq \prod_{l=1}^n \left(s^{\Psi}_l\right)^{\alpha_l} ,
\end{equation}
where $s^{\Psi}_l$ are the stabilizers defined in Eq.~\eqref{eq:psi-stabilizers} and the $\alpha_i \in [\dloc]$.
This implies that, for a given Pauli string $M$ acting on $\ket{\Psi}$,
there are $\dloc^n-1$ other Pauli strings that perform exactly the same action on $\ket{\Psi}$, namely, since
\begin{equation}\label{eq:eq-rel-M}
  M\ket{\Psi}=M\,S(\alpha_1,\ldots,\alpha_n)\ket{\Psi}, \end{equation}
all $\tilde M(\alpha_1,\ldots,\alpha_n) \coloneqq M S(\alpha_1,\ldots,\alpha_n)$ act identically on $\ket{\Psi}$.
All such \emph{realizations} $\tilde M(\alpha_1,\ldots,\alpha_n)$ of a Pauli string form an equivalence class.

The elements of such an equivalence class act on different subsets of the sites.
For example, the operator $X \otimes \1 \otimes \dots \otimes \1$, when acting on an AME state generated from a generator matrix in standard form, can be \emph{pushed} to the second half of the system by inserting the operator $(s_1^\Psi)^\dagger$ as then
\begin{align}
& X \otimes \1 \dots \1 \ket{\Psi} = (X \otimes \1 \dots \1) (s_1^\Psi)^\dagger \ket{\Psi} \\
& = \1 \dots \1 \otimes \overbrace{ X^{-g_{1,\floor{n/2}+1}}\otimes\ldots\otimes X^{-g_{1,n}}}^{\ceil{\frac{n}{2}}} \ket{\Psi} .
\end{align}
For the EPR state $\Phi^+ \coloneqq \sum_j \ket{j}\ket{j}$ this property is well known.
For every unitary $U$ acting on site 1 there is another unitary that transforms this state in the same way but acts only on site 2, i.e., $(U \otimes \1) \ket{\Phi^+} = (\1 \otimes U^T) \ket{\Phi^+}$.
In the case of AME states, any Pauli string can always be pushed to act non-trivially on at most any subset of size $\ceil{n/2}$:
\begin{lemma}\label{lemma:allpaulistringscanbepushed}
  For any given set of at most $\ceil{n/2}$ sites, any Pauli string $M$ acting on an $\AME(n,\dloc)$ state constructed from an MDS code according to Eq.~\eqref{amemingenerator} has a realization that acts non-trivially only on sites in this set, i.e., it can be pushed to act on at most these sites.
\end{lemma}
\begin{IEEEproof}
  This is a direct consequence of the fact that the tensor of coefficients of a minimal support AME state is a perfect tensor.
  More explicitly, constructing a realization $\tilde M(\alpha_1,\ldots,\alpha_n)$ that acts trivially on $\floor{n/2}$ sites is equivalent to solving two systems of each $\floor{n/2}$ linear equations, one for the powers of the $X$ operators and one for the powers of the $Z$ operators, because 
  \begin{align}
    &   \tilde M(\alpha_1,\ldots,\alpha_n) \\
    & = M\,\left(\bigotimes_{m=1}^{n} X^{\sum_{l=1}^{\floor{n/2}} \alpha_l\,g_{l,m}}\right)\left( \bigotimes_{m=1}^{n} Z^{\sum_{l=\floor{n/2}+1}^n \alpha_l\,h_{l,m}} \right) . \nonumber
  \end{align}
  As any subset of up to $\floor{n/2}$ columns of the generator and any subset of up to $\ceil{n/2}$ columns of the parity check are linearly independent, this can always be done.
\end{IEEEproof}

When pushing Pauli strings around, as the example above demonstrates, their weight can change.
In particular, it can happen that after pushing a Pauli string into a certain set of sites, it doesn't actually act non-trivially on all sites in this set.
We define the minimal weight of an equivalence class of operators as the weight of the ``lightest'' element within the class.
When a given $M$ belongs to a class of minimal weight $w$, this means that it cannot be pushed into any subset of less than $w$ sites, i.e., it can not be compressed to have weight less than $w$.

For the sake of simplicity we now restrict to the case $n$ even.
In the following theorem we show how a Pauli string $M$, belonging to a class of minimal weight $w$, defines an AME state based QECC.
\begin{theorem}\label{thm:qeccsfrommoperators}
  Given an AME state $\ket\Psi \in \AME(n,\dloc)$ constructed from an MDS code via Eq.~\eqref{amemingenerator} with $n$ even and $\dloc\geq n-1$ prime.
  The subspace $\C \coloneqq \lspan(\{\ket{\Psi_m}_{m=0}^{\dloc-1}\}) \subset (\mathbb{C}^\dloc)^{\otimes n}$, with
  \begin{equation}
    \ket{\Psi_m} \coloneqq M^m \ket{\Psi} 
  \end{equation}
  and $M$ a Pauli string, is a QECC spanned by AME states with code parameters $[\![n,1,w]\!]_\dloc$ if and only if $M$ belongs to an equivalence class of minimal weight $w \leq n/2$.
  Moreover, generators for the stabilizers group of the code state space can be constructed explicitly.
\end{theorem}
\begin{IEEEproof}
  The subspace is manifestly spanned by orthogonal AME states, as the $\ket{\Psi_m}$ are part of an orthonormal basis of AME states.
  This is a direct consequence of the fact that, because of Lemma~\ref{lemma:allpaulistringscanbepushed}, for $n$ even, a Pauli string acting on an AME state either stabilizes it or produces an orthogonal state, i.e., $\bra{\Psi} (M \ket{\Psi}) \in \{0,1\}$.
  Further, $\mathcal{C}$ is a QECC that can correct all errors from a set $\mathcal{E}$ if and only if there exists some function $f$ such that for all $E,F \in \mathcal{E}$ and all $m,m' \in [\dloc]$ \cite{Gottesman}
  \begin{equation}\label{eq:QECCcondition}
    \bra{\Psi_{m'}} E^\dagger\,F \ket{\Psi_{m'+m \mod \dloc}} = \delta_{m,0}\, f(E^\dagger\,F) .
  \end{equation}
  In our case, $\mathcal{E}$ is the set of all operators with weight at most $\floor{(w-1)/2} \leq \floor{(n-2)/4}$.
  We first prove the ``only if'' part.
  Assume that $M$ could be compressed into some subsystem of size less than or equal to $w-1$.
  As the compressed $M$ would still be a Pauli string and therefore product, there would be some error operators $E,F$ such that $E^\dagger\,F = M$ and hence the above condition would be violated.
  This proves necessity.
  We now turn to the ``if'' part.
  If $m=0$, then because $\ket{\Psi_{m'}}$ is an AME state
  \begin{equation}
    \bra{\Psi_{m'}} E^\dagger\,F \ket{\Psi_{m'+m \mod \dloc}} = \dloc^{n/2}\,\Tr(E^\dagger\,F) .
  \end{equation}
  Consider now the case $m \neq 0$.
  As $n$ is even, we know that we can push any Pauli string into any subset of size $n/2$.
  Denote the result of pushing $M^{m}$ with the product of stabilizers $S$ into some subset of size $n/2$ that completely contains the support of $E^\dagger\,F$ by $\widetilde{M^m} \coloneqq M^m\,S$.
  As $\ket{\Psi_{m'}}$ is AME
  \begin{align}
 \bra{\Psi_{m'}} E^\dagger\,F \ket{\Psi_{m'+m \mod \dloc}} &= \bra{\Psi_{m'}} E^\dagger\,F\,\widetilde M^{m}\ket{\Psi_{m'}} \\
& = \Tr(E^\dagger\,F\,\widetilde{M^{m}})\,\dloc^{n/2} .
  \end{align}
  Notice that $\widetilde{M^{m}}$ is not the $m$-th power of $M$ pushed in to the same subset, but $|\widetilde{M^m}| = |M^m\,S| = |(M\,S^{1/m})^m| = |M\,S^{1/m}|$ (as the Pauli matrices commute up to a phase, which does not change the weight) and so $M^m$ can be compressed into a certain number of sites if and only if $M$ can be compressed into the same number of sites.
  Thus $\widetilde{M^{m}}$ is, up to possibly a phase, a product of traceless Pauli operators that acts non-trivially on at least one site on which $E^\dagger\,F$ does not act, and therefore $\Tr(E^\dagger\,F\,\widetilde{M^{m}}) = 0$.
  This proves Eq.~\eqref{eq:QECCcondition} with $f(\cdot) = \dloc^{n/2}\,\Tr(\cdot)$.
  
  It remains to show how to construct the generators $s^{\mathcal{C}}_r$ of the stabilizers group of $\mathcal{C} = \lspan(\{\ket{\Psi_m}\}_{m=0}^{\dloc-1})$.
  The generators have to satisfy
  \begin{equation}
    \forall r,m\colon\quad s^{\mathcal{C}}_r\,M^m\,\ket{\Psi} = M^m\,\ket{\Psi} .
  \end{equation}
  The special case $m=0$ of this condition implies that they must be products of the stabilizers of $\ket\Psi$, i.e., that there exist vectors $\vec{\alpha}_r \in [\dloc]^n$ such that $s^{\mathcal{C}}_r = \prod_{l=1}^n (s^\Psi_l)^{(\vec{\alpha}_r)_l}$, hence they are in particular also Pauli strings.
  Further, the above condition implies that the $s^{\mathcal{C}}_r$ must commute with $M$ (and hence $M^m$) when acting on $\ket\Psi$, i.e,
  \begin{equation}\label{eq:mscommuteonpsi}
    s^{\mathcal{C}}_r\,M^m \ket\Psi = M^m\,s^{\mathcal{C}}_r \ket\Psi .
  \end{equation}
  The commutator of two Pauli strings, however is, up to a phase again a Pauli string.
  More precisely, any Pauli string $A$ can be brought into the standard form
  \begin{equation}
    A = \e^{\iu\,\varphi(A)}\,A_X^{\vec{a}^X}\,A_Z^{\vec{a}^Z} ,
  \end{equation}
  where $\varphi(A)$ is a phase, and $A_X^{\vec{a}^X} = \bigotimes_{j=1}^n X^{(\vec{a}^X)_j}$ and equivalently for $A_Z^{\vec{a}^Z}$.
  As can be verified by direct computation, for any two Pauli strings $A,B$ it holds that
  \begin{equation}
    A\,B = \omega^{\vec{a} \odot \vec{b}}\,B\,A
  \end{equation}
  where $\vec{a} = (\vec{a}^X,\vec{a}^Z)$ and $\vec{b}$ is defined in the same way in terms of the standard form of $B$ and $\odot$ is the bilinear symplectic inner product
  \begin{equation}
    \vec{a} \odot \vec{b} \coloneqq \vec{a}^Z \cdot \vec{b}^X - \vec{a}^X \cdot \vec{b}^Z .
  \end{equation}
  This implies that Eq.~\eqref{eq:mscommuteonpsi} is satisfied if for all $m$ it holds that $\vec{s}_r \odot m\,\vec{m} \mod \dloc = 0$, which is equivalent to just $\vec{s_r} \odot \vec{m} \mod \dloc  = 0$, where $\vec{s}_r$ is the vector coming from the standard representation of $s^{\mathcal{C}}_r$ and $\vec{m}$ that of $M$.
  More specifically
  \begin{equation}
    \vec{s}_r \coloneqq
    \begin{pmatrix}
      G^T & 0 \\
      0 & H^T
    \end{pmatrix}
    \vec{\alpha}_r 
  \end{equation}
  and as $M$ has weight less than $n/2+1$ the condition $\vec{s_r} \odot \vec{m} = 0$ imposes a non-trivial constrain, so that there are $n-1$ linearly independent $\vec{\alpha}_r$ that satisfy it and are of the form given above.
  A Mathematica code to find all these $\vec{\alpha}_r$ is given under \cite{amegithub}.
\end{IEEEproof}

In order for the code resulting from the above construction with a given Pauli string $M$ to be a QMDS code, the minimal weight $w$ of the Pauli strings equivalence class has to satisfy
\begin{equation}
  w = \floor{\frac{n-1}{2}+1} = \ceil{\frac{n}{2}}, 
\end{equation}
which precisely matches the lower bound set by Lemma~\ref{lemma:allpaulistringscanbepushed}.
We conjecture that for any $n$, and any AME states constructed in the above way from an MDS code, an equivalence class of Pauli strings exists that satisfies this bound (for $n$ even):  
\begin{conjecture}
  Given an AME state produced by an MDS code $\ket{\Psi}$ of $n$ sites and $\dloc\geq n-1$ prime, there exists at least one equivalence class of Pauli strings with minimal weight $\floor{n/2}$
\end{conjecture}
As a side remark, note that our proof of Theorem~\ref{thm:qeccsfrommoperators} in any case only works for $n$ even.

We have not been able to proof the above conjecture for all even $n$ but, using the computer algebra system Mathematica \cite{amegithub} we were able to construct incompressible $M$ operators with the conjectured properties for all $n \in \{2,4,6,7,8\}$, where $n=8$ is the largest $n$ for which we can exhaustively check all ways of pushing.
In all cases we were able to find such $M$ operators for $\dloc$ down to $\dloc = n-1$, except in the case $n=7$, where we had to chose $\dloc = 7$, because $n-1=6$ is not a power of a prime, and the case $n=2$, where our conjecture is known to be true for $\dloc \geq 2$ and the case $\dloc = 1$ does not make sense.
For $n=6$, the existence of the extended Singleton array $S_4'$ (see Eq.~\eqref{eq:extendedlingletionarray}) for $\dloc=4$ might give one hope that in this case an incompressible $M$ might exist for $\dloc = n-2$, but our calculations show that no such $M$ exists.
We summarize our results in Table~\ref{table:numericalresults}.

\begin{table}[h]
  \caption{List of QECCs whose existence we have verified by symbolic computation and exemplary $M$ matrices that generate a code form the respective family.
    All codes for $n$ even are QMDS, moreover the code $[\![7,1,3]\!]_\dloc$ is able to correct the same amount of errors as a QMDS code $[\![7,1,4]\!]_\dloc$.
    We do not obtain the codes $[\![5,1,3]\!]_{4,5,7,8}$ from our construction, but we have found $M$ operators not compressible to less than $d=3$ sites.}
\label{table:numericalresults}
\begin{center}
\begin{tabular}{l|l|l}
    QECC & $q$ & $M$ (first primitive element and smallest $q$)\\
    \hline
    $[\![3,1,1]\!]_\dloc$ & $2,3,4,5$ & $\1\otimes \1\otimes Z$\\
    $[\![4,1,2]\!]_\dloc$ & $3,4,5,7$ & $\1\otimes \1\otimes X \otimes Z$\\
    $[\![5,1,2]\!]_\dloc$ & $4,5,7,8$ & $\1\otimes \1\otimes \1 \otimes X \otimes Z$\\
    $[\![6,1,3]\!]_\dloc$ & $5,7,8,9,11,13$ & $\1\otimes \1\otimes X \otimes Z \otimes \1 \otimes Z$\\
    $[\![7,1,3]\!]_\dloc$ & $7,8$ & $Z \otimes \1\otimes Z \otimes \1\otimes \1\otimes \1\otimes Z$\\
    $[\![8,1,4]\!]_\dloc$ & $7,8$ & $\1\otimes \1\otimes \1\otimes Z \otimes \1\otimes Z \otimes Z \otimes X$
\end{tabular}
\end{center}
\end{table}

Further we can prove that Pauli strings containing only $X$ or only $Z$ operators and that have weight $\sim n/4$ can not be compressed to have weight less than $\sim n/4$:
\begin{lemma}
  Any Pauli string $M$ of weight $\wt(M) = \floor{(\floor{n/2}+1)/2}$ and consisting of only $X$ or only $Z$ operators is incompressible.
\end{lemma}
\begin{IEEEproof}
  Any product $S$ of stabilizers is again a stabilizer.
  Any such product $S$, apart from the trivial stabilizer, the identity, hence necessarily has weight at least $\wt(S) \geq \floor{n/2}+1$ (if it contains at least one $Z$ stabilizer, it actually has weight at least $\ceil{n/2} + 1$).
  If the weight of $M$ is $\wt(M) = \floor{(\floor{n/2}+1)/2}$, then for any $S$ we have $\wt(M\,S) \geq \min(\wt(M), (\floor{n/2}+1) - \wt(M)) = \wt(M)$.
\end{IEEEproof}
The above lemma shows that both the $X$ part $M_X$ and the $Z$ part $M_Z$ each consisting of $\floor{(\floor{n/2}+1)/2}$ many $X$ and $Z$ operators of an operator $M = M_X\,M_Z$ are individually incompressible.
If $M_X$ and $M_Z$ act on two disjoint sets of sites, and $n$ is even, then $\wt(M) \geq n/2$.
Unfortunately, the above lemma is not enough to guarantee that such an $M$ is also incompressible as a whole.
When pushing it into some subset of sites, its $X$ and $Z$ part can in principle start to overlap and thereby its weight can shrink. 
Our numerical calculations show that $M$ operators exist for which this does not happen up to the largest $n$ that we can check.

There is an interesting connection between incompressibility of $M$ matrices that contain both $X$ and $Z$ operators and a concept known as the \emph{joint weight enumerator} of two non-linear codes that are derived from the code generated by the $G_{\floor{n/2}\times n}$ matrix and its dual code with generator matrix $H_{\ceil{n/2}\times n}$. 
The joint weight enumerator $\mathscr{J}_{\mathscr{A},\mathscr{B}}$ of two classical codes $\mathscr{A}$ and $\mathscr{B}$ is a function of four real variables that encodes information about the overlap of zeros in the code words of the two codes \cite{Macwilliams1972,Rains1998} (see also \cite[Chapter 5]{MacWilliams}.
To relate this to the situation at hand, let $\mathscr{C}$ be the code generated by the $G_{\floor{n/2}\times n}$ matrix and $\mathscr{C}^{\perp}$ its dual code with generator matrix $H_{\ceil{n/2}\times n}$.
Now let $\mathscr{A}$ be the non-linear MDS code constructed from $\mathscr{C}$ by adding to each code word the vector of exponents of the $X$ operators in $M$ and $\mathscr{B}$ the non-linear MDS code constructed by adding to each code word of $\mathscr{C}^{\perp}$ the vector of exponents of the $Z$ operators.
Then the maximum number $i_{\mathrm{max}}$ of positions in which both a codeword from $\mathscr{A}$ and a codeword $\mathscr{B}$ have zeros is given by
\begin{equation}
  i_{\mathrm{max}} = \lim_{a \to \infty} \log \mathscr{J}_{\mathscr{A},\mathscr{B}}(a,1,1,1) .
\end{equation}
The minimal weight of the class of operators equivalent to $M$ is given by $n - i_{\mathrm{max}}$.

Let us return to the QECC we construct and compare its properties to known QECCs. 
Many other QECC for various combinations of parameters are known (see for example \cite{Ketkar2005,Grassl2015} for an overview and \cite{GrasslTable} for tables of known codes with $q=2$).
In some cases the achievable distances are limited by $\dC \leq q$ or some fraction of $\dloc$ \cite[Theorem 5]{Sarvepalli2005} or scale only like $\sqrt{n}$ \cite[Theorem 40 and 41]{Ketkar2005}.
In general, it is difficult to construct QECC that saturate the quantum Singleton bound and have a large code distances \cite{Jin2013}.
For example in \cite[Corollary 32]{Ketkar2005} a QMDS code with a code distance of the order of $n/2$ was shown to exist, the proof however requires that $\dloc$ grows faster than exponentially with $n$.
Families of QMDS codes with a code distance up to $\dC = \dloc + 1$ have been constructed in \cite{Jin2013,Grassl2015} (for an earlier construction with $\dC = \dloc$ see \cite{Grassl2003}), but all of these (as well as the code from \cite[Theorem 5]{Sarvepalli2005} when $\dC$ is chosen to scale linear with $\dloc$) have code lengths $n$ that scale quadratically with $\dloc$.

Our construction only requires $n \leq \dloc +1$ to achieve a code distance that scales like $n/2$ (or equivalently like $\dloc/2$).
We can show the existence of these codes with $n = \dloc +1$ explicitly for $n = \{4,6,8\}$, and if our above conjectuer holds true, an infinite family of such codes for arbitrarily large and even $n$ exists.
For physical implementations with independent local noise it appears to be important to achieve a large ratio $\dC/n$.
In this respect our codes perform well compared with the constructions discussed above.

There are two construction that are similar to ours in terms of code parameters.
The first was presented in \cite{Aharonov1996} and later used in \cite{Cleve1999}.
It yields QECCs for all prime $\dloc$ with $\dloc > n = 2\,\dC-1$.
The second is based on \cite[Lemma~70]{Ketkar2005}, which shows that the existence of a pure stabilizer QECC $[\![n,k,\dC]\!]_\dloc$ with $n,\dC \geq 2$ implies the existence of a code $[\![n-1,k+1,\dC-1]\!]_\dloc$.
The possibility to construct stabilizers for AME states with $\dloc$ prime (see section Section~\ref{sec:stabilizers}) and the fact that such AME states are QECCs of the form $[\![\dloc+1,0,\floor{(\dloc+1)/2}+1]\!]_\dloc$ \cite[Proposition~3]{Scott} implies the existence of QECCs of the form $[\![\dloc,1, \floor{(\dloc+1)/2}]\!]_{\dloc}$ for all $\dloc$ prime.
The first construction requires $\dloc \geq n+1$ and the second works in the case $\dloc = n$, but neither of the the two can straightforwardly be expanded to the case $\dloc = n-1$.
To sum up, as a function of $\dloc$, there are constructions that achieve larger code distances $\dC$ and larger $k$ than our proposal, but all such constructions we are aware of require (asymptotically) larger code lengths $n$.

\section{Conclusions}
In this paper we have shown in detail how to explicitly construct AME states of $n$ parties with local dimension $\dloc \geq n -1$ of minimal support by means of linear MDS codes.
For an AME state of minimal support constructed via such a linear MDS code and $\dloc$ prime, 
we have derived a set of stabilizer operators that stabilize the AME state.
For every $n \leq \dloc+1$ we show how to construct QECCs that encode a logical qudit into a $\dloc$-dimensional subspace spanned by AME states of $n$ parties with local dimension $\dloc$ prime.
Under a conjecture for which we provide numerical evidence, this construction produces an infinite family of quantum error correcting codes for arbitrary large $n$ that, for $n$ even, saturate the quantum Singleton bound.
Along the way, we have also shown how, starting from any single AME state, a complete basis of the Hilbert space consisting of AME states can be constructed.

\section*{APPENDIX}\label{app:construction-Singleton-AMEstate}
As a first example let us consider the case $\dloc=2^2$.
The elements of the finite field $GF(2^2)$ can be written in several different ways (see Table~\ref{table:GF(4)}).
As the field is a non-prime finite field, it will be convenient to work with the representation in terms of polynomials based on the irreducible polynomial $x^2=x+1$.

\begin{table}[hb]
\caption{$GF(2^2)$ generated by $x^2=x+1$.}
\label{table:GF(4)}
\begin{center}
\begin{tabular}{ccc}
\hline 
as a 2-tuple & as a polynomial & spin levels \\ 
\hline 
00 & 0 & 0 \\ 
10 & 1 & 1 \\ 
01 & x & 2 \\ 
11 & x+1 & 3 \\ 
\hline 
\end{tabular}
\end{center}
\end{table}
As we have seen in \eqref{eq:extendedlingletionarray}, for the special case $q=2^2$ we can use the extended singleton array
\begin{equation} \label{eq:extendedlingletionarrayappendix}
S_4' = \begin{array}{cccc}
1 & 1 & 1 & 1\\
1 & a_1 & a_2 &   \\
1 & a_2 & a_1  &   \\
1 &   &  &  \\ 
 \end{array}.
\end{equation}
We chose $\gamma = x$ as a primitive element and, using the polynomial representation of $GF(2^2)$, the appearing elements can be calculated to be
\begin{align}
a_1 &= \frac{1}{1-x}= x \\
a_2 &= \frac{1}{1-x^2}=\frac{1}{x}=x+1 .
\end{align}
Let us now take the the biggest submatrix of size $3\times 3$ as the $A$ matrix 
\begin{equation}
  A = \left[\begin{array}{ccc}
1 & 1 & 1 \\
1 & x & x+1 \\
1 & x+1 & x  \\
 \end{array} \right]
\end{equation}
and construct an AME state in $\AME(6,4)$.
The number of free indices in the closed form expression of a state in $\AME(6,4)$ with minimal support is $k = 3$, so  $\vec v = (i, j, l)$, and it can be written as,
\begin{align} \label{AME6,4}
  & \AME(6,4) \ni \ket\Psi = \\
  & \sum_{\substack{i,j,l \in\\ \{0,1,x,x+1\} }}  \ket{i,j,l,i+j+l,i+x\,j+(1+x)\,l,i+(x+1)\,j+x\,l } \nonumber
\end{align}	
To get the terms of $\ket\Psi$ in the familiar computational basis representation, after doing all computations in the finite field for each term, one simply has to switch back to the spin level representation, i.e., make the replacement $\{0,1,x,x+1\} \mapsto \{0,1,2,3\}$ according to Table~\ref{table:GF(4)}. 

We can now also find generators of the stabilizer group:
\begin{align}
s_1 &= X\otimes\1\otimes\1\  \otimes   X\otimes X^{\phantom{2}}\otimes X^{\phantom{2}}\\
s_2 &= \1\otimes X\otimes\1\  \otimes  X\otimes X^2\otimes X^3\\
s_3 &= \1\otimes \1\otimes X\ \otimes  X\otimes X^3\otimes X^2\\
\cr
s_4 &= Z^3\otimes Z^3\otimes Z^3\  \otimes  Z\otimes \1\otimes \1\\
s_5 &= Z^3\otimes Z^2\otimes Z^{\phantom{2}}\  \otimes  \1\otimes Z\otimes \1\\
s_6 &= Z^3\otimes Z^{\phantom{2}} \otimes Z^2\  \otimes  \1\otimes \1\otimes Z
\end{align}
All the elements of stabilizer group $S$, which are $\dloc^n=5^6$ elements, can be generated from all possible product  combinations of the stabilizer generators.

Finally, we present a number of Singleton arrays that can be used to construct closed form expression of AME states with minimal support in Table~\ref{table:Singleton arrays}. 
A Mathematica notebook to create these and various larger tables is made available under \cite{amegithub}.

\begin{table*}[hb]
    \caption{Singleton array for various finite fields.}
    \label{table:Singleton arrays}
    \hfill
    \begin{minipage}{0.9\textwidth}
\begin{center}
\begin{tabular}{ |c |c @{, } c|}
\hline 
\begin{tabular}{c}\\$GF(2)=\{0,1\}$\\\text{modulo (2)} \end{tabular}  & 
$\gamma=1$ &
$S_2=\begin{array}{cc}1 & 1  \\ 1 &  \end{array}$ 
\\ 
\hline
\begin{tabular}{c}\\$GF(3)=\{0,1,2\}$\\\text{modulo (3)} \end{tabular} & 
$\gamma=2$ &
$S_3=\begin{array}{ccc}1 & 1 & 1 \\ 1 & 2 & \\ 1 & & \end{array}$\\  
\hline
\begin{tabular}{c}\\$GF(2^2)=\{0,1,a_1,a_2\}$\\\text{modulo ($1+x+x^2$)} \\\ \\ $a_1=x$, $a_2=1+x.$\end{tabular}& 
$\gamma=x$ &
$S'_4=\begin{array}{cccc}1 & 1 & 1 & 1 \\ 1 & a_1 & a_2 & \\ 1 & a_2 & a_1 & \\ 1& & & \end{array}$  \\ 
\hline
\begin{tabular}{c}$GF(5)=\{0,1,2,3,4\}$\\\text{modulo (5)}\end{tabular} &
$\gamma=3$ &
$S_5=\begin{array}{ccccc}1 & 1 & 1 & 1 & 1\\ 1 & 2 & 3 & 4 & \\ 1 & 3 & 4 & & \\ 1& 4 & & & \\ 1 & & & &\end{array}$  \\  
\hline
\begin{tabular}{c}$GF(7)=\{0,1,2,3,4,5,6\}$\\\text{modulo (7)}\end{tabular}  &
$\gamma=3$ &
$S_7=\begin{array}{ccccccc}1 & 1 & 1 & 1 & 1 & 1 & 1\\ 1 & 3 & 6 & 4 & 2 & 5 & \\ 1 & 6 & 4 & 2 & 5 & & \\ 1& 4 & 2 & 5 & & &\\1& 2 & 5 & & & &\\ 1 & 5 & & & & & \\ 1 & & & & & \end{array}$   \\ 
\hline
\begin{tabular}{c}$GF(2^3)=\{0,1,a_1,a_2,a_3,a_4,a_5,a_6\}$\\ \text{modulo ($1+x^2+x^3$)}\\\  \\ $a_1=x^2,\ a_2=1+x+x^2,\ a_3=1+x,$ \\$ a_4=x,\ a_5=x+x^2$ and $a_6=1+x^2.$\end{tabular}&
$\gamma=x$ &
$S_8=\begin{array}{cccccccc} 1 & 1 & 1 & 1 & 1 & 1 & 1 & 1 \\ 1 & a_1& a_2 & a_3 & a_4 & a_5 & a_6 & \\1 & a_2 & a_3 & a_4 & a_5 & a_6 & & \\1 & a_3 & a_4 & a_5 & a_6 & & & \\1 & a_4 & a_5 & a_6 & & & &\\1  & a_5 & a_6 & & & & &\\ 1  & a_6 &  & & & & & \\ 1 & & & & & & & \end{array}$ \\
\hline
\begin{tabular}{c}$GF(3^2)=\{0,1,a_1,a_2,a_3,a_4,a_5,a_6,a_7\}$\\ \text{modulo ($2+x+x^2$)}\\\ \\ $a_1=2+x,\ a_2=1+x,\ a_3=1+2x,$ \\$ a_4=2,\ a_5=x,\ a_6=2x$ and $a_7=2+2x.$\end{tabular}&
$\gamma=x$ &
$S_9=\begin{array}{ccccccccc} 1 & 1 & 1 & 1 & 1 & 1 & 1 & 1 &1 \\ 1 & a_1& a_2 & a_3 & a_4 & a_5 & a_6 & a_7 \\1 & a_2 & a_3 & a_4 & a_5 & a_6 & a_7 & & \\1 & a_3 & a_4 & a_5 & a_6 & a_7 & & & \\1 & a_4 & a_5 & a_6 & a_7 & & & &\\1  & a_5 & a_6 & a_7 & & & & &\\ 1  & a_6 & a_7 & & & & & & \\ 1 & a_7 &  & & & & & & \\ 1 & & & & & & & & \end{array}$ \\
\hline
\begin{tabular}{c}$GF(11)=\{0,1,2,3,4,5,6,7,8,9,10\}$\\\text{modulo (11)}\end{tabular}  & 
$\gamma=2$ &
$S_{11}=\begin{array}{ccccccccccc}1 & 1 & 1 & 1 & 1 & 1 & 1 & 1 & 1 & 1 & 1 \\ 1 & 10 & 7 & 3 & 8 & 6 & 4 & 9 & 5 & 2 & \\ 1  & 7 & 3 & 8 & 6 & 4 & 9 & 5 & 2 & & \\ 1  & 3 & 8 & 6 & 4 & 9 & 5 & 2 & & & \\  1  & 8 & 6 & 4 & 9 & 5 & 2 & & & & \\ 1  & 6 & 4 & 9 & 5 & 2 & & & & & \\ 1 & 4 & 9 & 5 & 2 & & & & & & \\ 1& 9 & 5 & 2 & & & & & & & \\  1 & 5 & 2 & & & & & & & & \\  1 & 2 & & & & & & & & & \\  1 & & & & & & & & & &  \end{array}$ \\
[1cm]\hline 
\end{tabular}
\end{center}
    \end{minipage}\hfill
\vspace*{8ex}
\end{table*}

\section*{Acknowledgments}

We would thank Sara Di Martino, Dardo Goyeneche, Markus Johansson, Jos\'e Ignacio Latorre, Marco Piani, German Sierra and Karol \.Zyczkowski for discussions as well as Markus Grassl, Daniel Gottesman, and Felix Huber for very useful comments on an earlier version of this manuscript. 
We acknowledge support from the European Research Council
(CoG QITBOX), Axa Chair in Quantum Information Science, 
Spanish MINECO (QIBEQI FIS2016-80773-P, FISICATEAMO FIS2016-79508-P and Severo Ochoa Grant No.~SEV-2015-0522),
Fundaci\'{o} Privada Cellex, and Generalitat de Catalunya (Grant No.~SGR 874, 875, and CERCA Programme). 
C.\ G.~acknowledges support by MPQ-ICFO, ICFOnest+ (FP7-PEOPLE-2013-COFUND), and co-funding by the European Union's Marie Sk\l{}odowska-Curie Individual Fellowships (IF-EF) programme under GA: 700140.
A.\ R.~is supported by the Beatriu de Pin\'os fellowship (BP-DGR 2013)

\end{document}